\documentclass[a4paper,11pt]{article}
\usepackage{pos}
\usepackage{lineno}
\usepackage{siunitx}
\usepackage{multirow}
\usepackage{amsmath}
\usepackage{tablefootnote}
\title{Performance of the New FlashCam-based Camera in the 28\,m Telescope of H.E.S.S.}

\ShortTitle{FlashCam Performance on H.E.S.S.-CT5}
        
\author*[a]{Baiyang~Bi}
\author[b]{Miquel~Barcelo} 
\author[b]{Christian~Bauer}
\author[b]{Faical~Ait~Benkhali}
\author[c]{Jacqueline~Catalano}
\author[c]{Sebastian~Diebold}
\author[b]{Christian~F\"ohr}
\author[c]{Stefan~Funk}
\author[d]{Gianluca~Giavitto}
\author[b]{German~Hermann}
\author[b]{Jim~Hinton}
\author[c]{Ira~Jung-Richardt}
\author[b]{Oleg~Kalekin}
\author[b]{Ruben~Kankanyan}
\author[b]{Thomas~Kihm}
\author[a]{Fabian~Leuschner}
\author[c]{Marc~Pfeifer}
\author[a]{Gerd~P\"uhlhofer}
\author[e]{Olaf~Reimer}
\author[b]{Simon~Sailer}
\author[a]{Heiko~Salzmann}
\author[a]{Andrea~Santangelo}
\author[b]{Simon~Steinmassl}
\author[a]{Thomas~Schanz}
\author[a]{Chris~Tenzer}
\author[b]{Felix~Werner}

\affiliation[a]{Institut f\"ur Astronomie und Astrophysik, Eberhard Karls Universit\"at T\"ubingen, Sand 1, 72076 T\"ubingen, Germany}
\affiliation[b]{Max-Planck-Institut f\"ur Kernphysik, PO Box 103980, 69029 Heidelberg, Germany}
\affiliation[c]{Physikalisches Institut, Friedrich-Alexander Universit\"at Erlangen-N\"urnberg, Erwin-Rommel-Str. 1, D 91058 Erlangen, Germany}
\affiliation[d]{DESY, D-15738 Zeuthen, Germany}
\affiliation[e]{Institut f\"ur Astro- und Teilchenphysik, Leopold Franzens Universt\"at Innsbruck, Technikerstrasse 25/8, A 6020 Innsbruck, Austria}
        
\emailAdd{baiyang.bi@astro.uni-tuebingen.de}

\abstract{In October 2019, the central \SI{28}{\m} telescope of the H.E.S.S. experiment has been upgraded with a new camera. The camera is based on the FlashCam design which has been developed in view of a possible future implementation in the Medium-Sized Telescopes of the Cherenkov Telescope Array (CTA), with emphasis on cost and performance optimization and on reliability. The fully digital design of the trigger and readout system makes it possible to operate the camera at high event rates and to precisely adjust and understand the trigger system. The novel design of the front-end electronics achieves a dynamic range of over 3,000 photoelectrons with only one electronics readout circuit per pixel. Here we report on the performance parameters of the camera obtained during the first year of operation in the field, including operational stability and optimization of calibration algorithms.}

\FullConference{37$^{\rm{th}}$ International Cosmic Ray Conference (ICRC 2021)\\
		July 12th -- 23rd, 2021\\
		Online -- Berlin, Germany}

\begin{document}
\maketitle
\section{Introduction}
The H.E.S.S. (High Energy Stereoscopic System) observatory is located in the Khomas Highland in Namibia. It consists of four \SI{12}{\m} and one \SI{28}{\m} imaging atmospheric Cherenkov telescopes.
The four \SI{12}{\m} telescopes (CT1-4) started operation in 2003, and the central \SI{28}{\m} telescope (CT5) in 2012 \cite{HESS2017}.
In October 2019, after seven years of operation, the original camera on CT5 was replaced with an advanced prototype of a FlashCam camera to improve the telescope performance and stability.\par
FlashCam has been developed as a camera candidate for the Medium-Sized Telescope (MST) of the Cherenkov Telescope Array (CTA) \cite{FlashCam2008,FlashCam2013,FlashCam2017,FlashCam2019}. 
Based on a fully-digital design of the readout and trigger system, FlashCam is able to run with zero dead time with a trigger rate of up to \SI{30}{\kilo\hertz}.
The FlashCam pre-amplifier saturates in a controlled way when the amplitude exceeds 3,000 times a least significant bit (LSB) value, entering a non-linear regime.
The non-linear pre-amplifier extends the dynamic range of the FlashCam signal chain by one order of magnitude, and therefore permits to digitize each pixel's signal with only one amplifier stage and readout channel.
After many years of prototyping and extensive simulations, the signal reconstruction algorithm of FlashCam has ultimately been verified with a full-size prototype 
at the Max-Planck-Institut f\"ur Kernphysik (MPIK), Heidelberg. \par 
An advanced version of the prototype was installed on H.E.S.S.-CT5 in October 2019, followed by a short commissioning and scientific verification phase.
The camera has been running stably for more than one and a half years by now. Here, we report on some key performance experiences that have been gained in the field with this new camera type.

\section{FlashCam}
\subsection{FlashCam Design}
The FlashCam core elements comprise the photon detection plane (PDP), the readout system (ROS), and the data acquisition system (DAQ).
The PDP detects the incident photons with photo-multiplier tubes (PMTs), and produces analog differential signals with pre-amplifiers.
There are 1758 active channels belonging to 147 modules on the PDP.
Winston cones are installed in front of each PMT to improve the photon collection efficiency and to restrict the sensitive angular range.
A UV-transparent window is installed in front of the PDP to protect PMTs and Winston cones.
A water cooling system keeps the median PDP temperature at roughly \SI{30}{\degreeCelsius}, which protects the electronics and minimizes gain changes of the PDP.
The analog signals are transmitted via CAT. 5E shielded twisted-pair cables to the ROS, 
where they are sampled at a rate of \SI{250}{\mega\hertz} with 12-bit flash analog-to-digital converters (fADCs).
The design of the ROS is based on a fully digital approach with continuous signal digitization, 
which runs dead time free with standard settings at a trigger rate of up to \SI{30}{\kilo\hertz}.
The sampled traces are stored in the buffer of FPGAs and processed to the trigger system to derive the trigger decisions. 
The camera is connected to a camera server via up to four 10-Gbit Ethernet fibres, transferring triggered data from the camera to the server.
The DAQ's camera server is located at a central computing cluster.
Two additional Gbit Ethernet fibres connect the camera server to the camera, for slow control and monitoring, and for providing the array-wide precision clock signal through a White Rabbit network, respectively.
More details of the FlashCam system have been given in \cite{FlashCam2017, FlashCam2019}.

\subsection{Lab Calibration}
\begin{figure}[htbp]
    \centering
    \includegraphics[scale= 0.28]{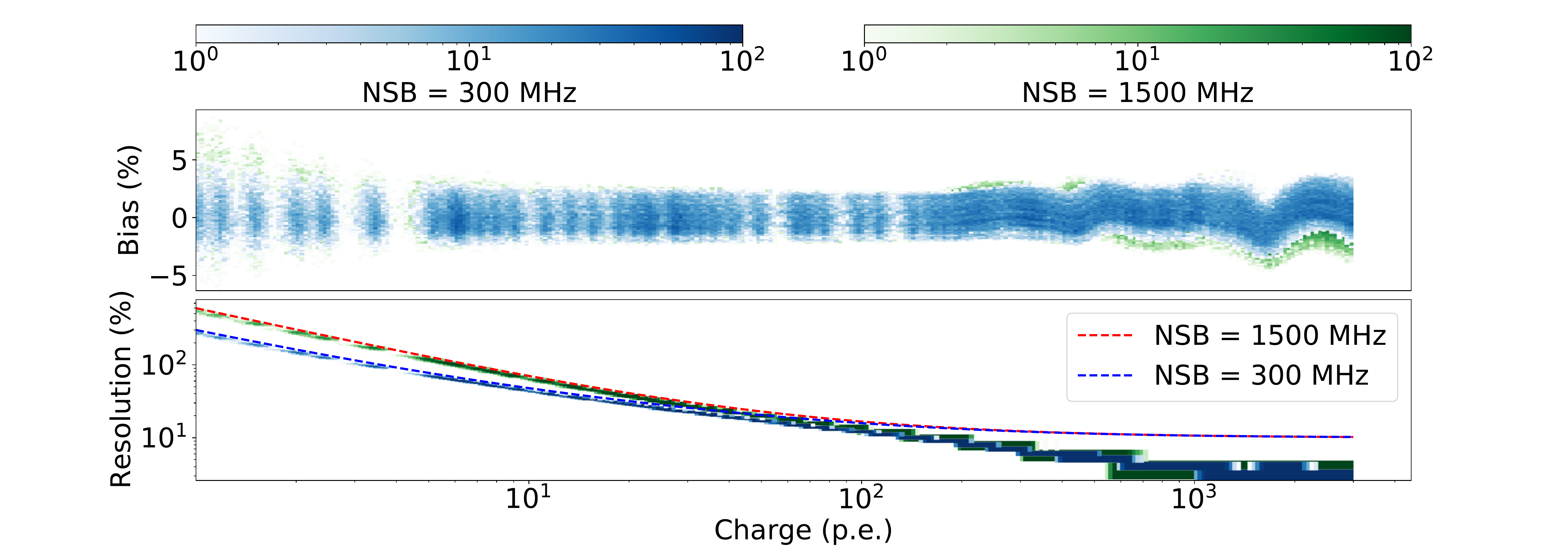}
    \caption{The lab calibration results of two NSB levels, i.e. \SI{300}{\mega\hertz} and \SI{1500}{\mega\hertz}.
    The upper plot is the bias of the reconstruction. 
    The lower plot is the resolution comparing to the CTA benchmark functions (the dash lines). 
    The color scale is the number of pixels in each bin.}
    \label{fig:Bias_Res}
\end{figure}

The calibration of the full-size prototype was completed at MPIK.
A laser with a wavelength of \SI{355}{\nm} is used as light source.
A neutral density filter wheel %with the optical density (OD) linearly changing from 0 to 4, 
is placed in front of the laser to simulate pulse intensities from 0.3 to a few thousand photoelectrons (p.e.) per pixel.
%\textcolor{red}{@Felix: I copied the number from your NIM-A publication, and confirmed it with the analysis. But, how can an OD-4 filter wheel can produce 0.3 to 10000 p.e.? 0.3*1e4=3000!!!}
A diffuser in front of the filter wheel spreads the light across the whole surface of the camera. 
A UV LED in DC mode is used to simulate the effect of the night sky background (NSB)
up to a rate of \SI{4.8}{\giga\hertz} per pixel, more details can be found in \cite{FlashCam2017}.
%To calibrate the intensity of the attenuated laser pulse in the non-linear regime, 
%three pixels are not equipped with Winston cones and three more pixels are partially covered, 
%in order to maintain linearity when unmasked pixels are saturated.
%They are referred to as masked pixels.
The linear and non-linear reconstruction algorithms have been well calibrated in the lab.
The bias and the charge resolution over the full dynamic range are shown in Fig. \ref{fig:Bias_Res}. 
The dynamic range reaches up to 3,000 p.e. with a maximal bias of roughly \SI{5}{\percent}
and the charge resolution complies with the CTA benchmark functions used during development of the camera \cite{FlashCam2017}.
%The CTA requirement is defined as: 

%\begin{align}
%     \sigma_\mathrm{Q} &= \sqrt{\sigma_0^2 + \sigma_{\mathrm{ENF}}^2Q + \sigma_\mathrm{g}^2Q^2}/Q \label{eq:resolution_req}\\
%%     \sigma_0 &= \sqrt{f_{\mathrm{NSB}}\cdot t_\mathrm{w} + n_\mathrm{e}^2} \\
%     \sigma_{\mathrm{ENF}} &= 1 + \mathrm{ENF}
%\end{align}
%where $\sigma_\mathrm{Q}$ is the charge resolution for the charge $Q$. $\sigma_0$ is an additive component, 
%including the contribution from the NSB ($f_{\mathrm{NSB}} \cdot t_\mathrm{w}$) 
%and from the electronics noise ($\sigma_e=0.87$).
%The NSB rate ($f_{\mathrm{NSB}}$) is determined from the shift of the baseline,
%and the contribution from NSB is the charge in the time window ($t_\mathrm{w}$) of 15 ns.  
%$\sigma_{\mathrm{ENF}}$ is determined by the excess noise factor (ENF), and $\sigma_\mathrm{g}$ by the uncertainty of the gains of PMTs, which gives the miscalibration factor. 
%$\mathrm{ENF}=0.2$ and $\sigma_\mathrm{g}=0.1$ are used to estimate the requirement.

\section{Performance data from H.E.S.S.-CT5}
\subsection{Installation into CT5}
In October 2019, an advanced FlashCam prototype was installed into CT5. 
The net weight load of the FlashCam system was adjusted to be the same as that of the previous camera.
That minimized the change of the mechanics of the telescope.
Because of the different size of the new and previous camera, 
the distance from the optical target to the camera entrance plane was slightly changed.
Thus, the focus settings were adjusted by sweeping the camera position 
to place the camera entrance plane at the image plane of the shower at 15 km distance in zenith,
which is the typical shower height scale.
Observations started in November 2019, and have been running stably more than one and a half years by now.
The camera has been available for data taking more than \SI{98}{\percent} of the time.
During this period not a single channel nor electronics board broke.
%\textcolor{blue}{German added this, if so, the three broken pixels should be deleted.}
The performances of sub-systems are reported in the following subsections.
\subsection{The Cooling System}
\begin{figure}[htbp]
    \centering
    \includegraphics[scale=0.45]{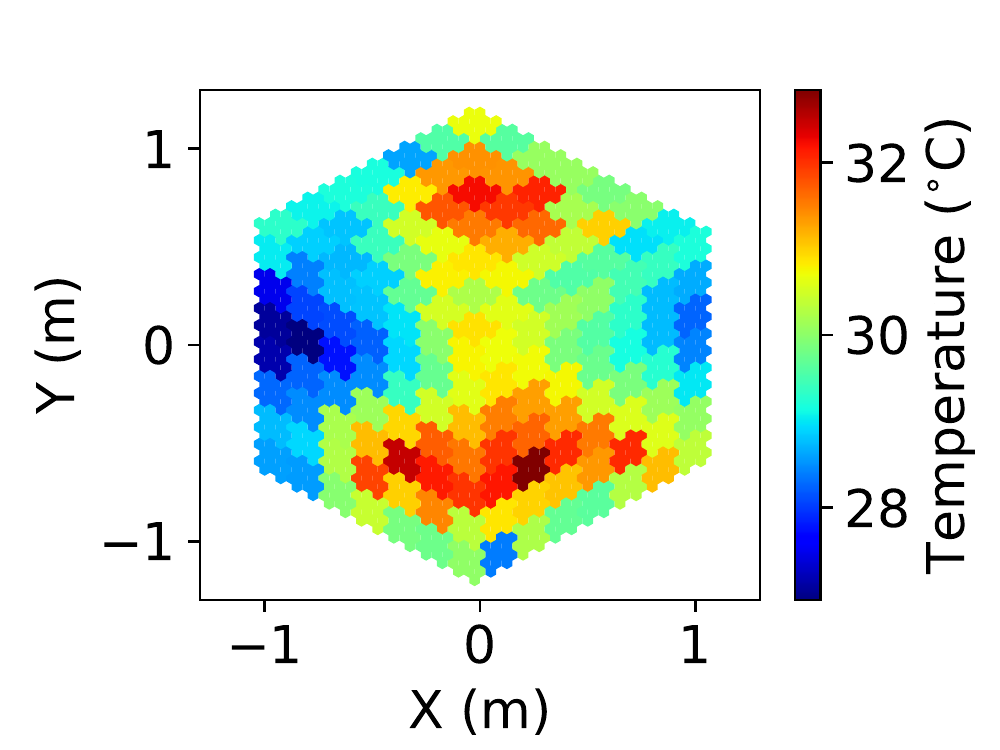}
    \includegraphics[scale=0.45]{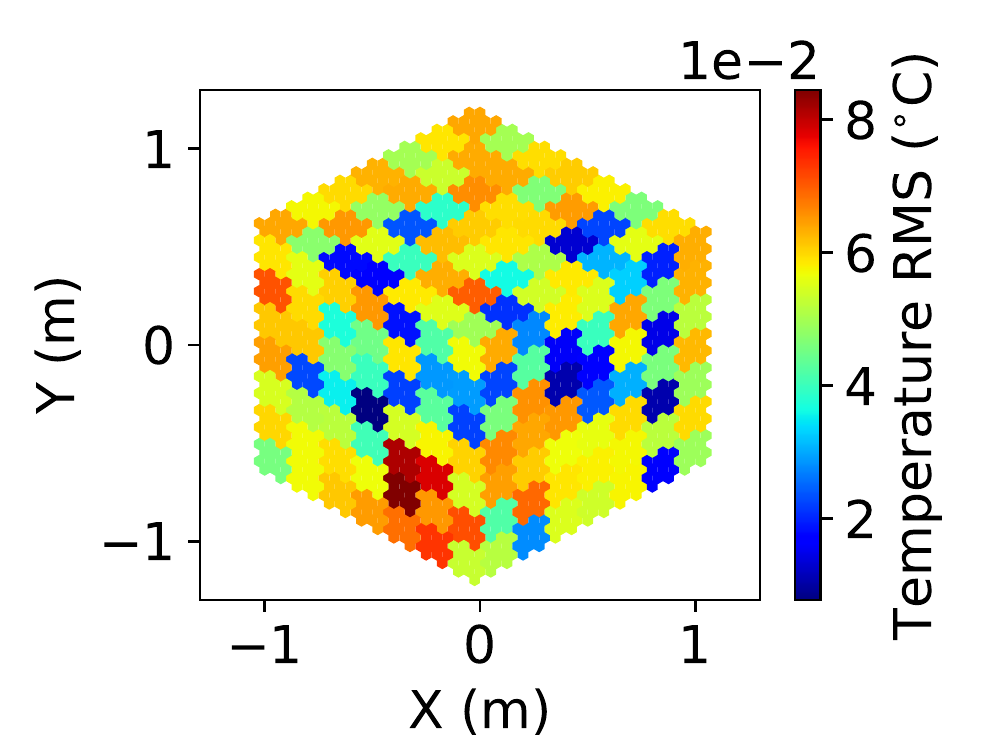}
    \includegraphics[scale=0.45]{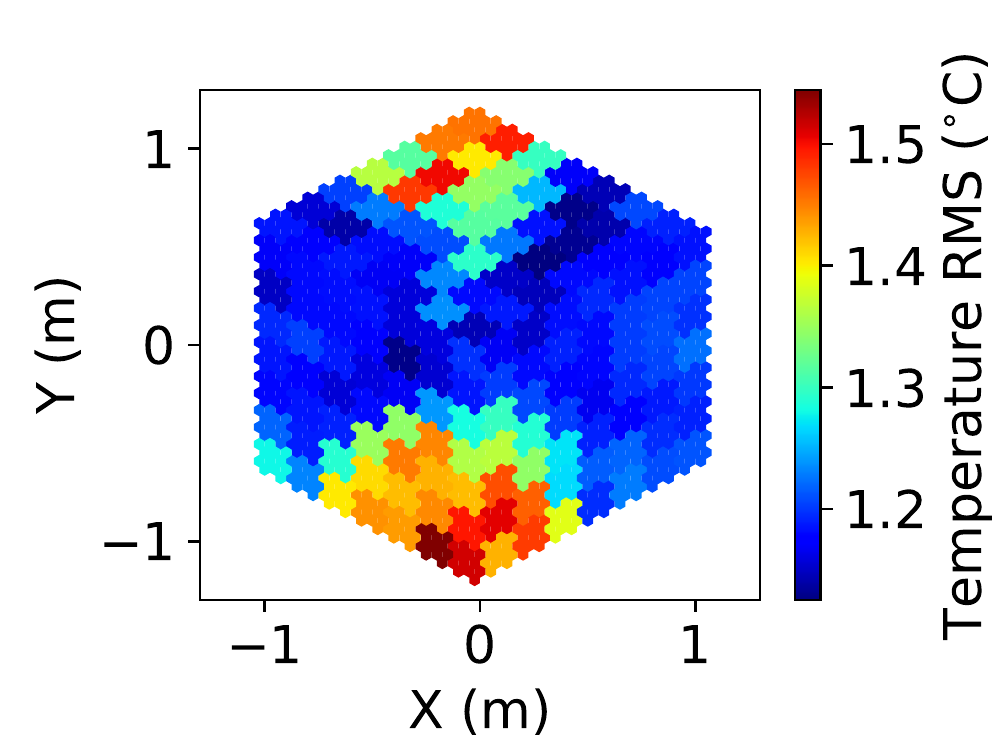}\\
    \includegraphics[scale=0.45]{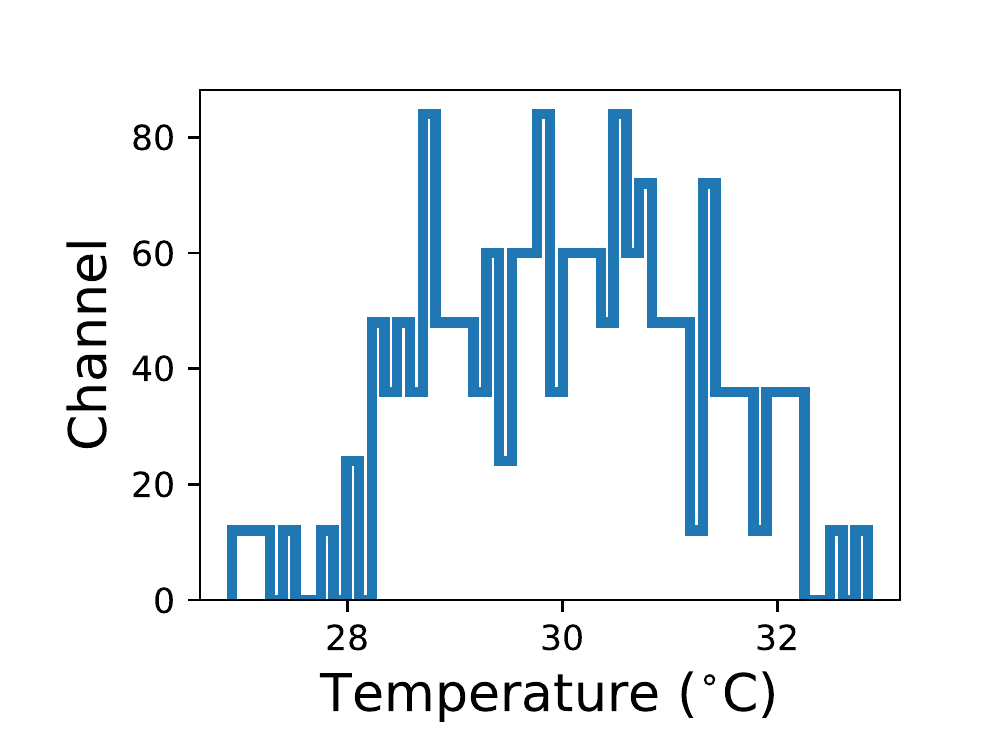}
    \includegraphics[scale=0.45]{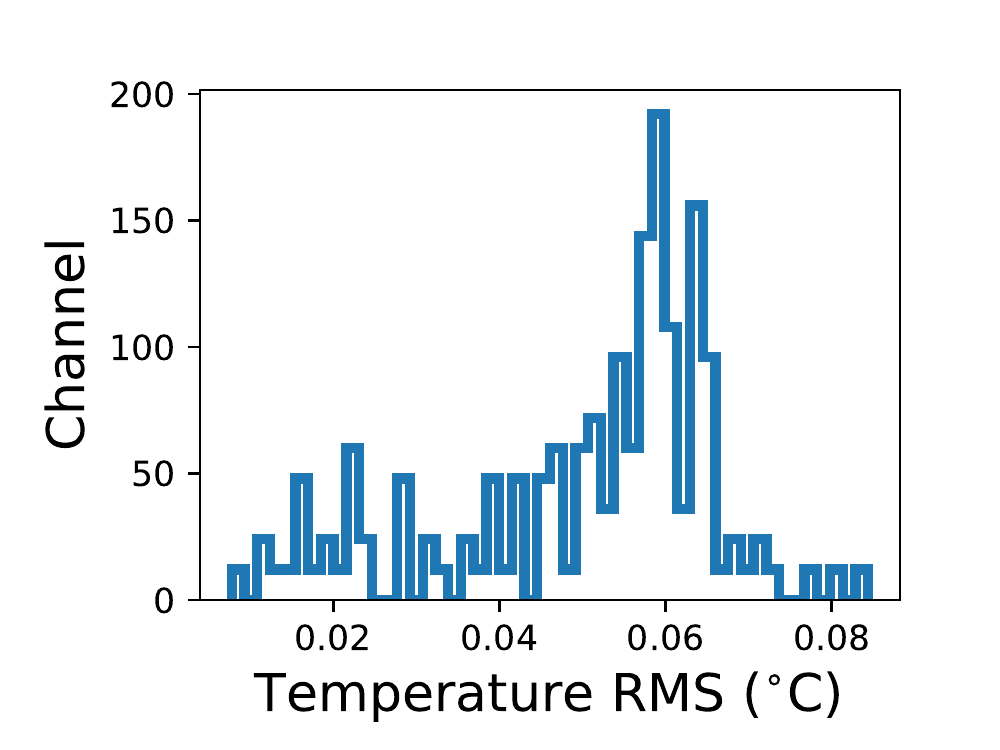}
    \includegraphics[scale=0.45]{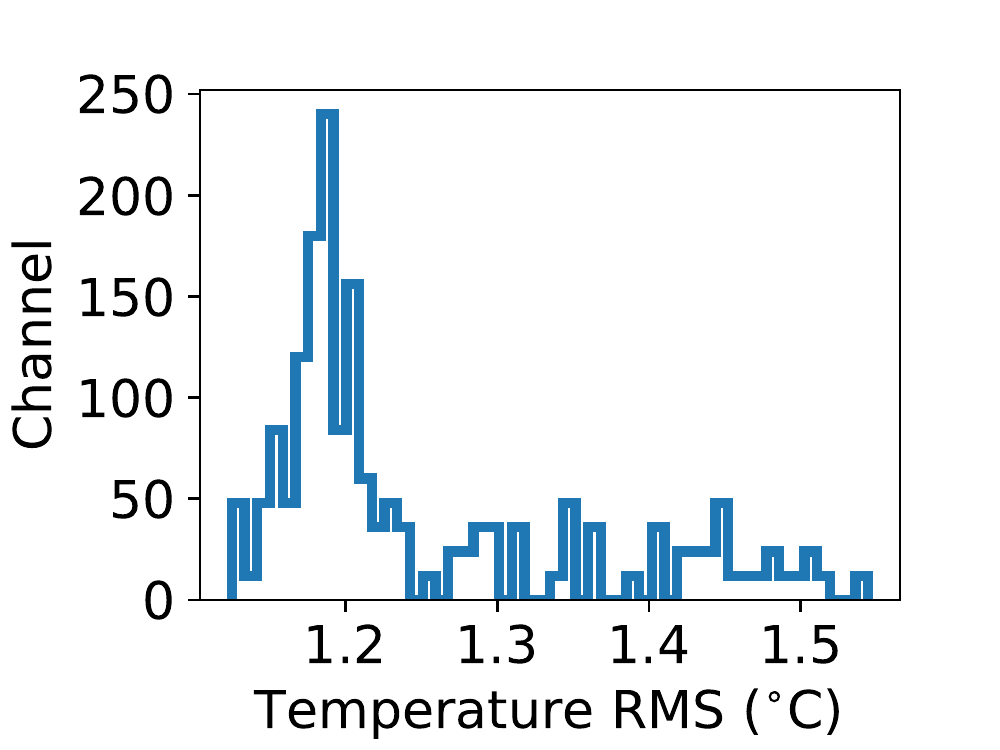}
    \caption{The PDP temperatures. Left: The temperature during one run (Run 160149). 
    The uneven distribution is due to the cooling water going into the camera from the left and right side 
    and leaving at the upper and lower side. Middle: The RMS of the PDP temperatures during run 160149.
    Right: The RMS of the PDP temperatures over one year of operation (2019-11 to 2020-12).}
    \label{fig:temperature}
\end{figure}

The camera housing and much of the tubing are thermally insulated. 
An active water cooling system is used to keep a sufficiently constant camera temperature during night-time observations.
The design of the cooling system with its tubing is more demanding in CT5 than in CTA-MSTs,  
due to the $\sim$ \SI{60}{\m} height variation during operation.
The optimal range for the median camera-internal temperature during observations is 26 to \SI{32}{\degreeCelsius}. 
Thanks to the cooling system, this can be maintained and the typical standard deviation of the temperature for a specific pixel is less than \SI{0.1}{\degreeCelsius} during a 28-minute run.
%Thanks to the cooling system, the optimal range for the median camera-internal temperature (the PDP temperature) 
%during observations is 26 – 32$^{\circ}$C, 
%with the typical standard deviation of the temperature of a specific pixel is less than 0.1$^{\circ}$C during a run.
The maximum temperature standard deviation of the pixels is roughly \SI{1.5}{\degreeCelsius} (see Fig. \ref{fig:temperature}). 

\subsection{Photo-detector modules}
\begin{figure}[htbp]
    \centering
    \includegraphics[scale=0.2]{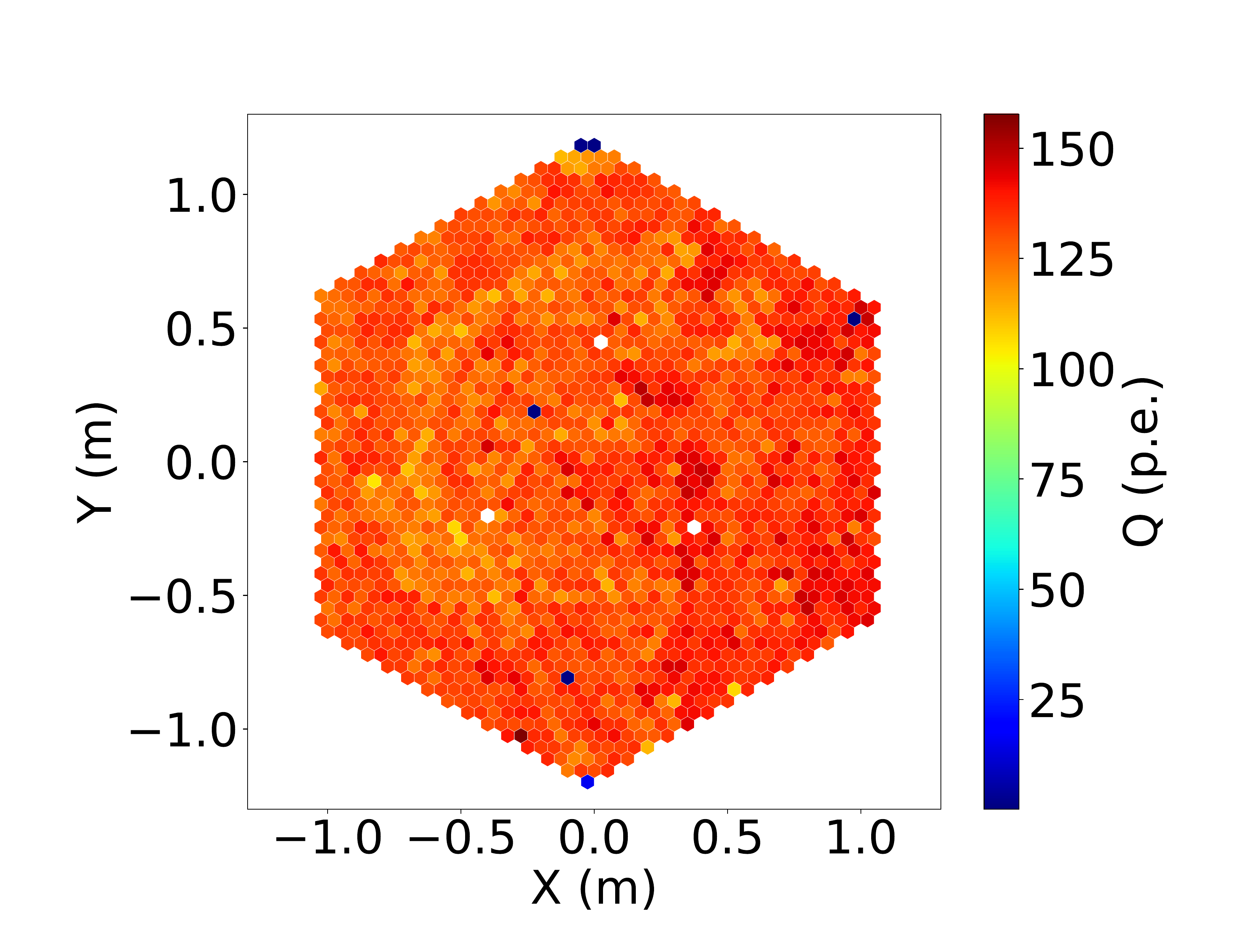}
    \includegraphics[scale=0.2]{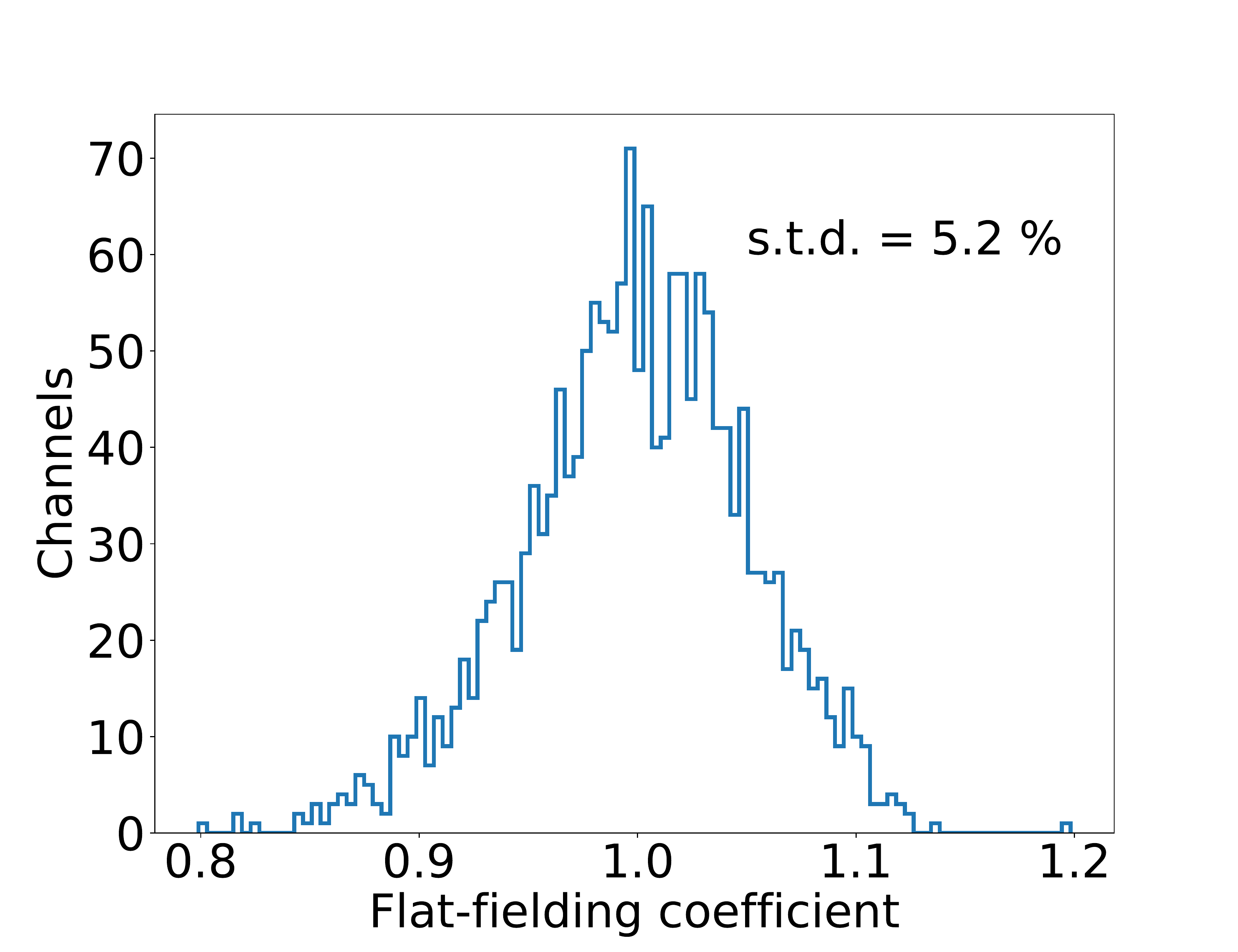}
    \caption{Left: The intensity of an example flat-fielding run (taken on 2020-04-11). Right: The distribution of the flat-fielding coefficients.}
    \label{fig:pe_CT5}
\end{figure}

\begin{figure}[htbp]
    \centering
    \includegraphics[scale=0.2]{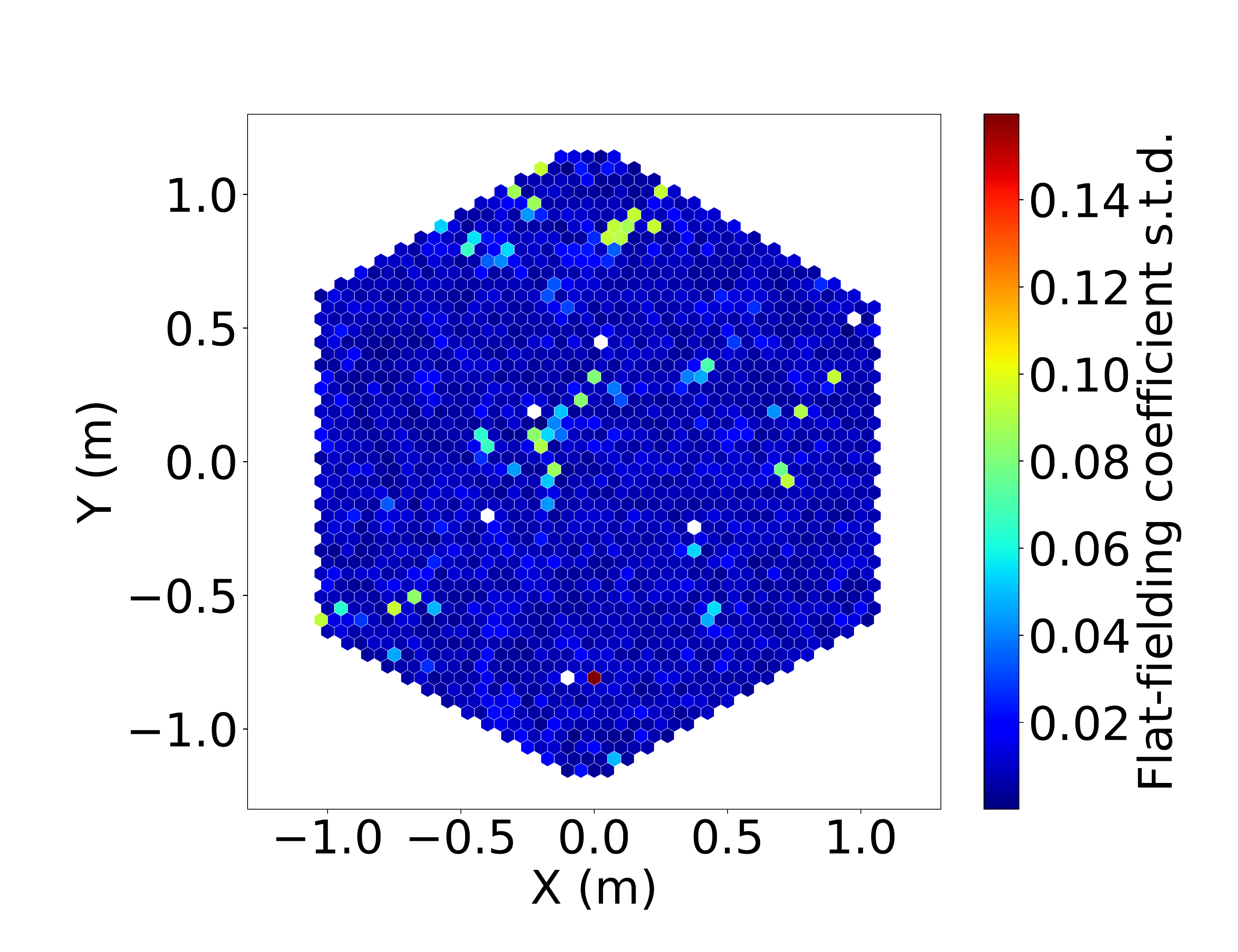}
    \includegraphics[scale=0.2]{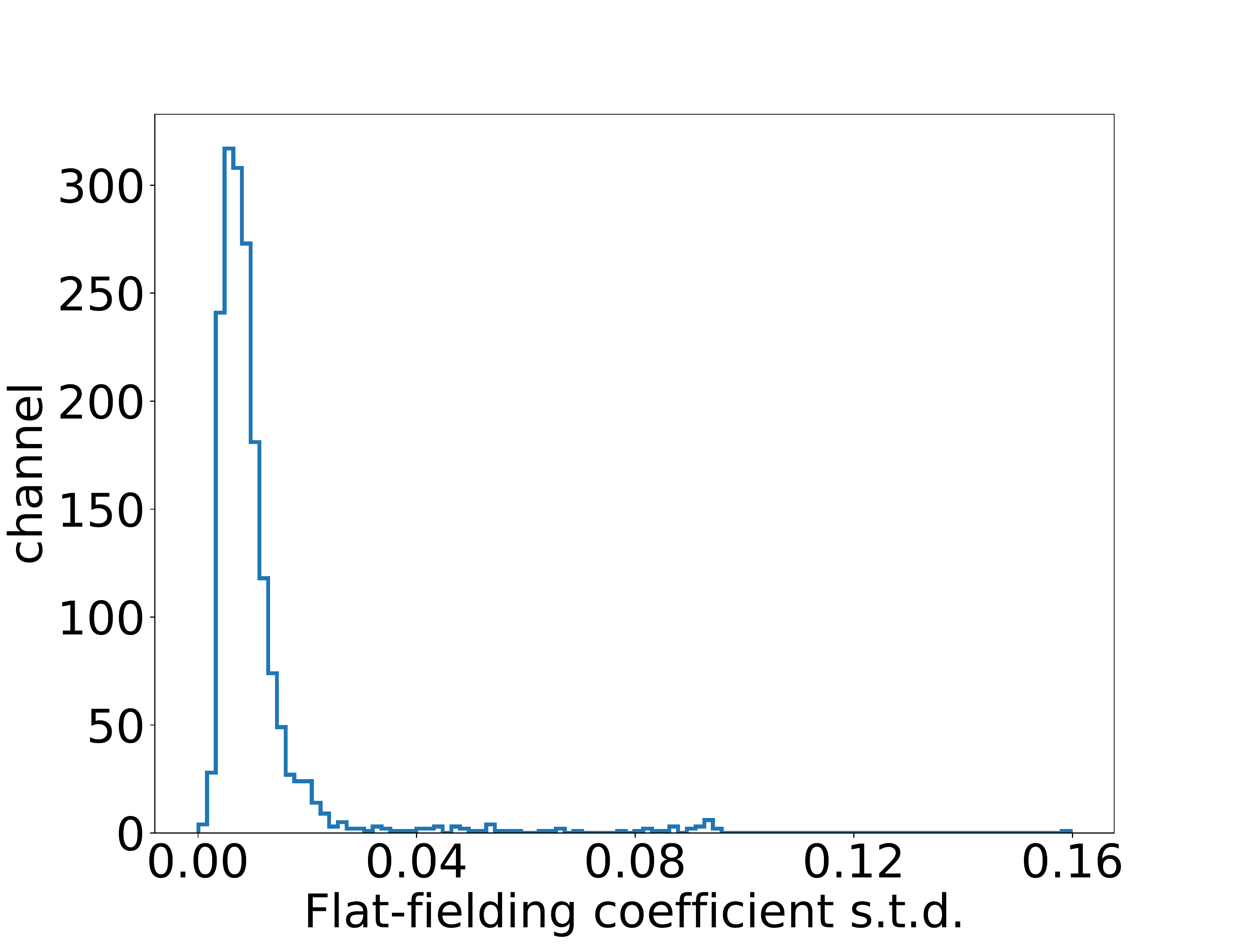}
    \caption{The standard deviation of the flat-fielding coefficients in one year (2019-11 to 2020-12), excluding the masked pixels.}
    \label{fig:ff_rms}
\end{figure}

\begin{figure}[htbp]
    \centering
    \includegraphics[scale=0.25]{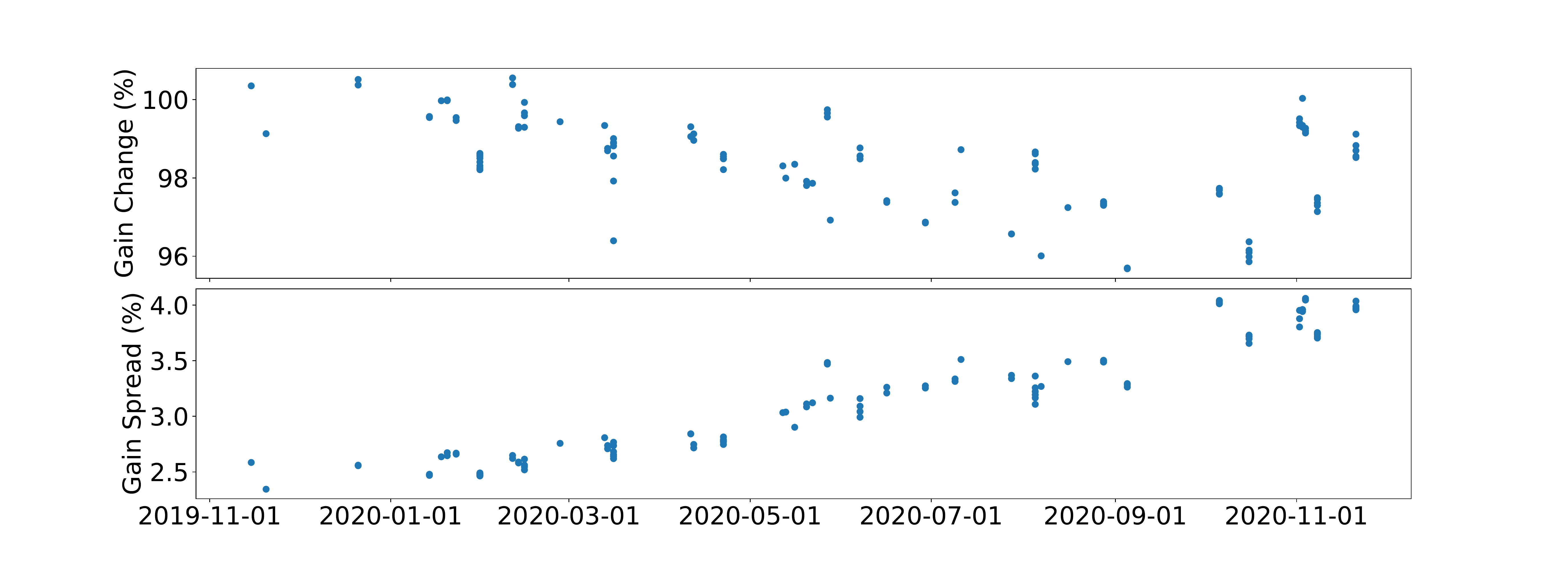}
    \caption{The gain change and gain spread between 2019-11 and 2020-12.}
    \label{fig:CT5Gain}
\end{figure}

The PDP was calibrated and monitored with a flat-fielding unit.
The flat-fielding unit is installed at the center of the mirror dish of CT5, 
roughly \SI{36}{\m} away from the camera.
The flat-fielding unit (FF-unit) consists of 13 LEDs and the voltage on these LEDs can be adjusted between 8 and 16 Volts, 
corresponding to an intensity of 100 p.e. up to a few thousand p.e.
It illuminates the camera nearly uniformly with a repetition rate of \SI{1}{\kilo\hertz}.
Fig. \ref{fig:pe_CT5} shows the the intensity of the FF-unit on the PDP with a median of roughly 130 p.e.
There are three pixels, two on the top and one on the bottom, 
which are partially masked to calibrate the intensity of the FF-unit  in the non-linear regime of the camera.
Three pixels equipped with struts instead of PMTs 
are used to support the UV-transparent protective window in front of the PDP 
(the three white pixels forming an equilateral triangle in Fig. \ref{fig:pe_CT5}).
Three further pixels marked in dark blue have been deactived at the start of operation.
%\textcolor{red}{Besides, there are another three pixels in darkblue which were broken after the initial commissioning.
%No attempts have been made to recover these three channels and no further channels broke since then.} 
%\textcolor{blue}{German would like to delete these two red sentences.}
Inpainting methods have been established in the analysis framework to successfully minimise the impact of
any deactived or missing pixel.
%\textcolor{red}{three broken pixels and the three window-supporting pixels.}
%\textcolor{blue}{if the broken pixels deleted, here should be "of these pixels"}
Fig. \ref{fig:pe_CT5} also shows the flat-fielding coefficients of the pixels of an example flat-fielding run.
The flat-fielding coefficient is defined as the ratio of the charge of each pixel
%(except for the supporting, masked and broken pixels) 
to the median charge over the camera.
The standard deviation of the flat-fielding coefficients of the pixels is roughly \SI{5.2}{\percent}.
The stability of the flat-fielding coefficients is monitored with the flat-fielding runs
that are performed in each month.
As shown in Fig. \ref{fig:ff_rms}, the variation of the flat-fielding coefficients of most of the pixels is less than roughly \SI{3}{\percent}.
The change of the dynode gains of the PMTs are monitored as well, using the FF unit events. 
Assuming a Poisson distribution of the injected charge and constant excess noise of the PMTs,
the single photoelectron (spe) responses of the pixels are derived based on the statistical properties of the light pulses from the CT5 flat-fielding unit. 
Monthly averages over the entire camera up to the end of 2020 are plotted in Fig. \ref{fig:CT5Gain}, 
and show a $\sim$ \SI{4}{\percent} gain loss with respect to that in November 2019, 
while the gain spread (the relative standard deviation) increased from \SI{2.5}{\percent} to \SI{4}{\percent}.\par
\begin{figure}[htbp]
    \centering
    \includegraphics[scale=0.35]{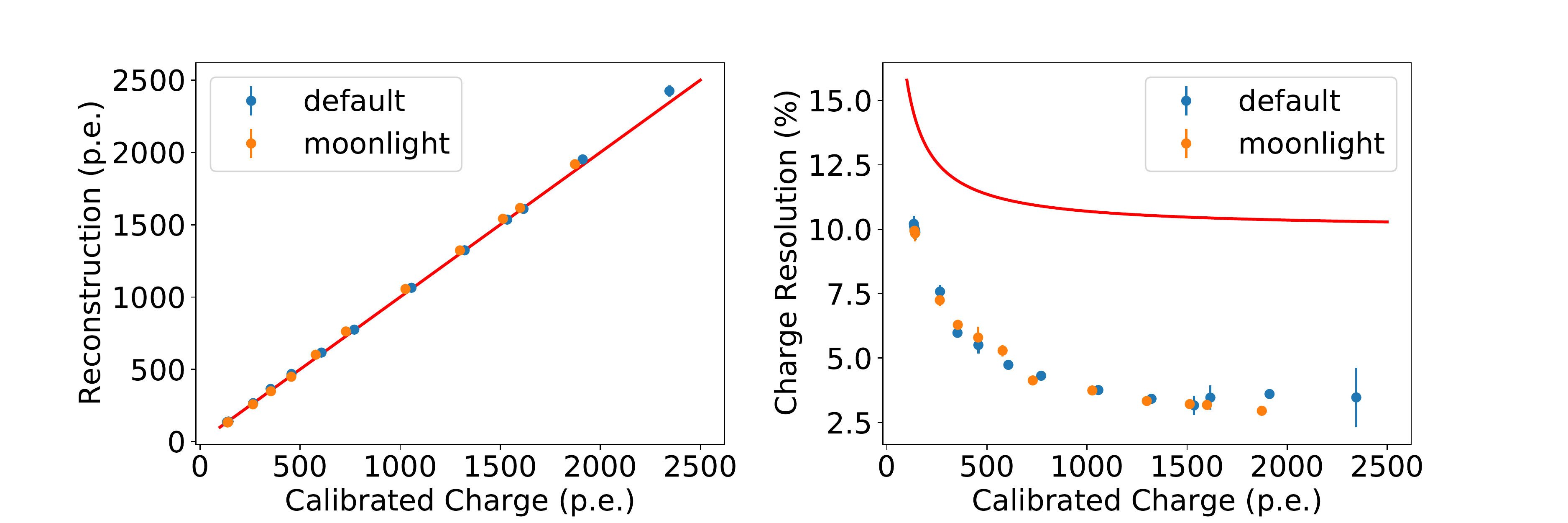}
    \caption{Results from the intensity sweep runs. Left: Reconstructed charge v.s. calibrated charge. The red line is the expectation, i.e. reconstructed charge equaling the calibrated charge. Right: The charge resolutions, compared to the CTA benchmark (the red line).}
    \label{fig:LEDSweep}
\end{figure}

At the end of January 2020, a set of intensity sweep runs were performed,
by sweeping different number of LEDs and LED intensities,
to study the performance of the FlashCam and to verify the reconstruction algorithm in the non-linear regime.
The performances with the default setup and moonlight setup\footnote{To mitigate the impact of the increased aging under moonlight conditions, 
the high voltages of the PMTs are lowered for the "MoonlightObservation" runs
reducing the gain of PMTs to half of the default values.}
were both verified, with intensities from 130 p.e. (the normal flat-fielding runs) up to more than 2000 p.e.
The results are shown in Fig. \ref{fig:LEDSweep}.
The reconstruction agrees with the calibration which was done with the masked pixels 
and the resolution is better than the CTA benchmark\footnote{The CTA benchmarks for different NSB levels don't differ much above 100 p.e. (see Fig. \ref{fig:Bias_Res}).}.
\subsection{Timing}
\begin{figure}[htbp]
    \centering
    \includegraphics[scale=0.45]{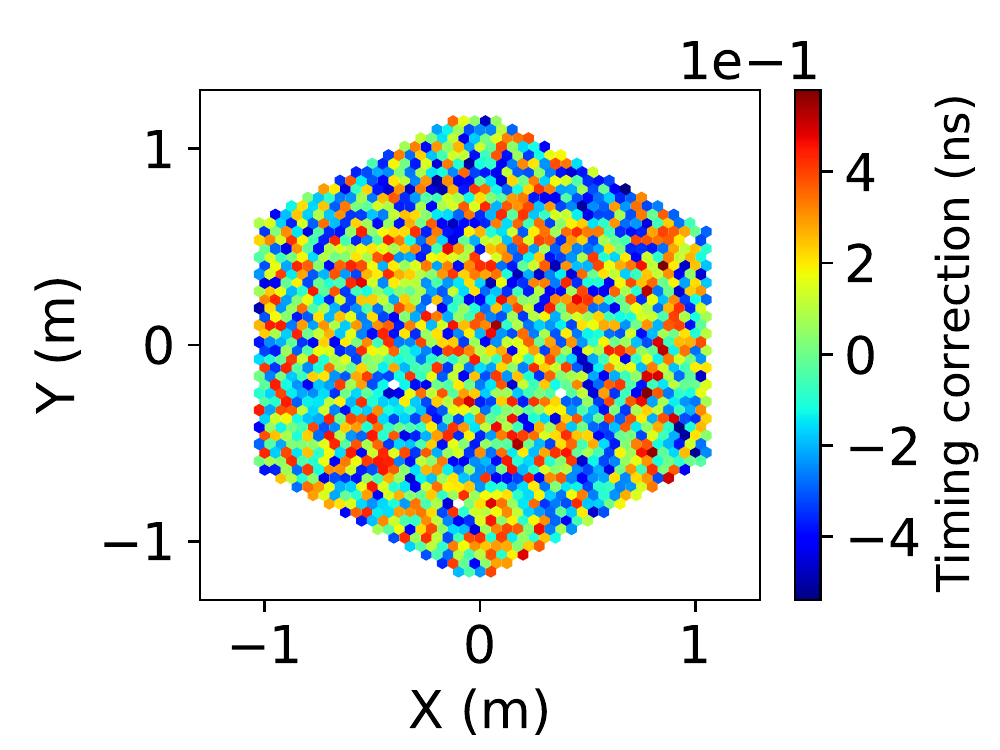}
    \includegraphics[scale=0.45]{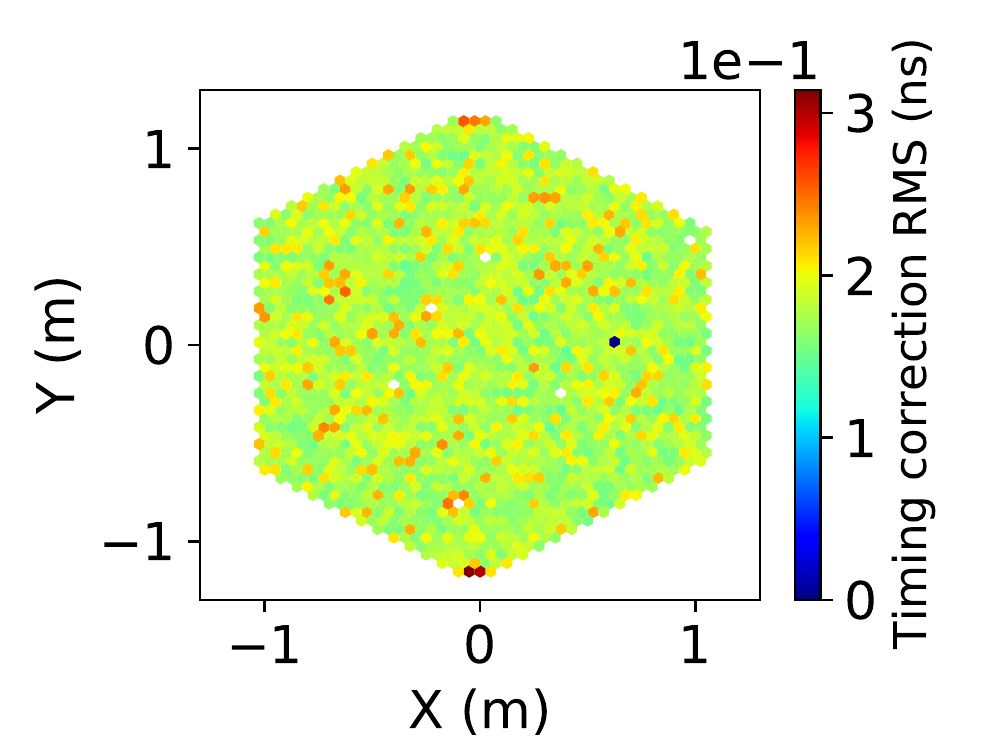}
    \includegraphics[scale=0.45]{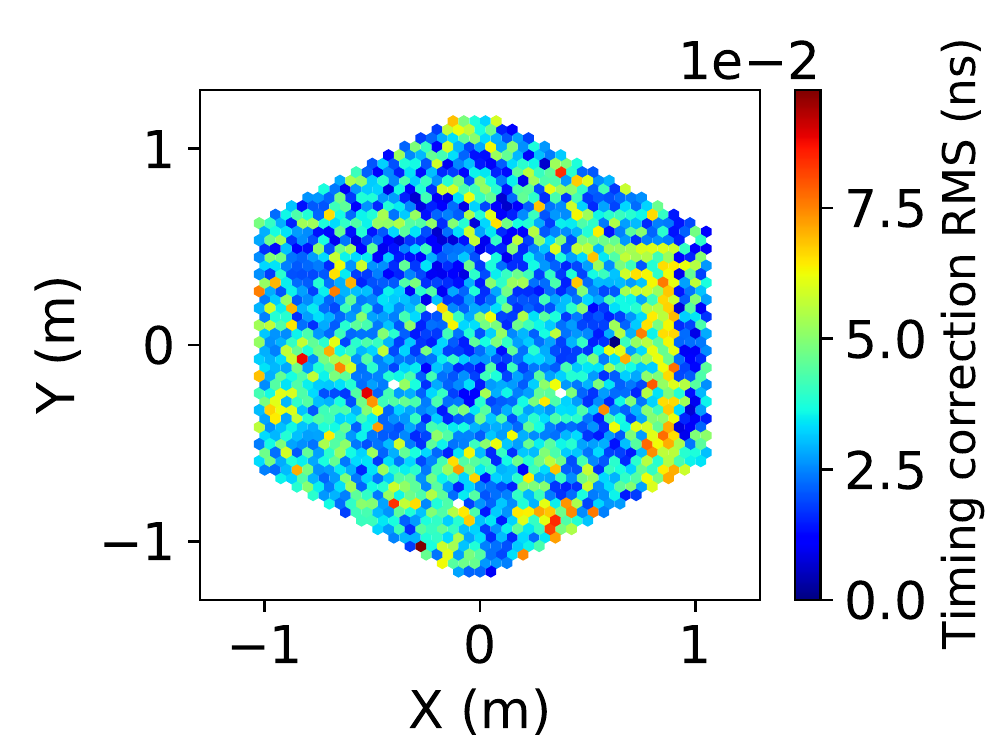}\\
    \includegraphics[scale=0.45]{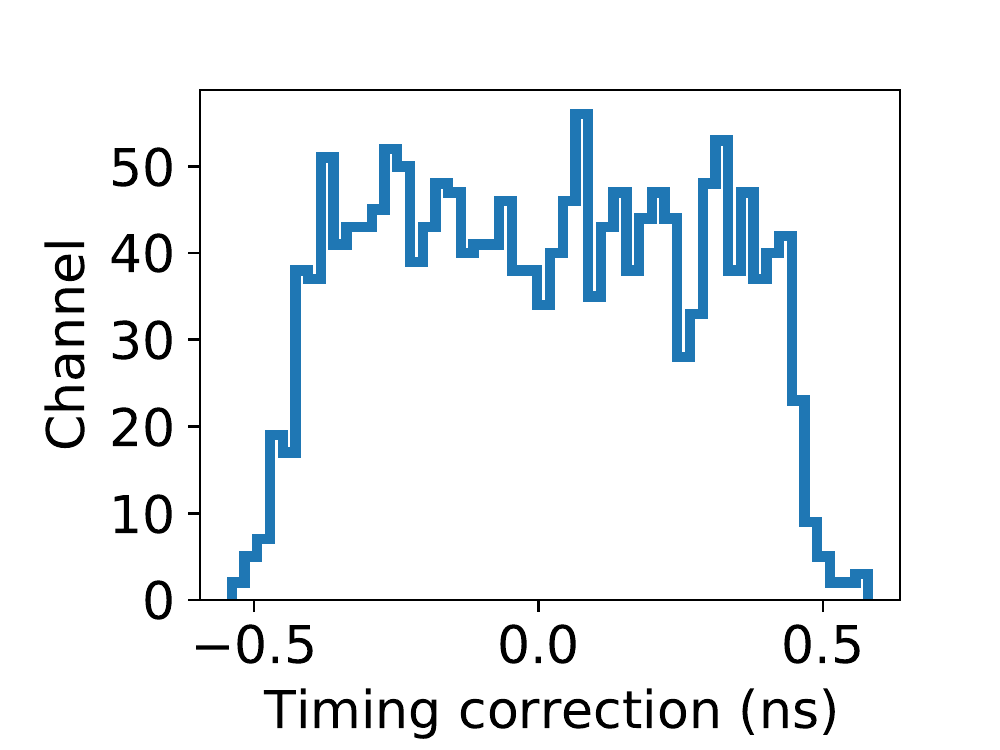}
    \includegraphics[scale=0.45]{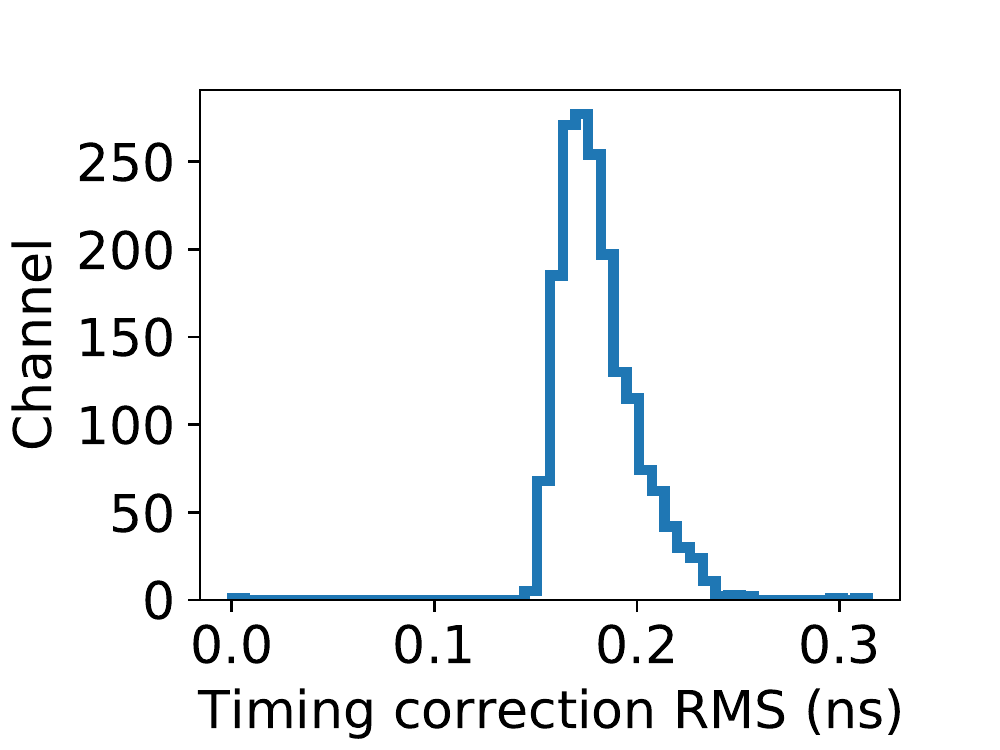}
    \includegraphics[scale=0.45]{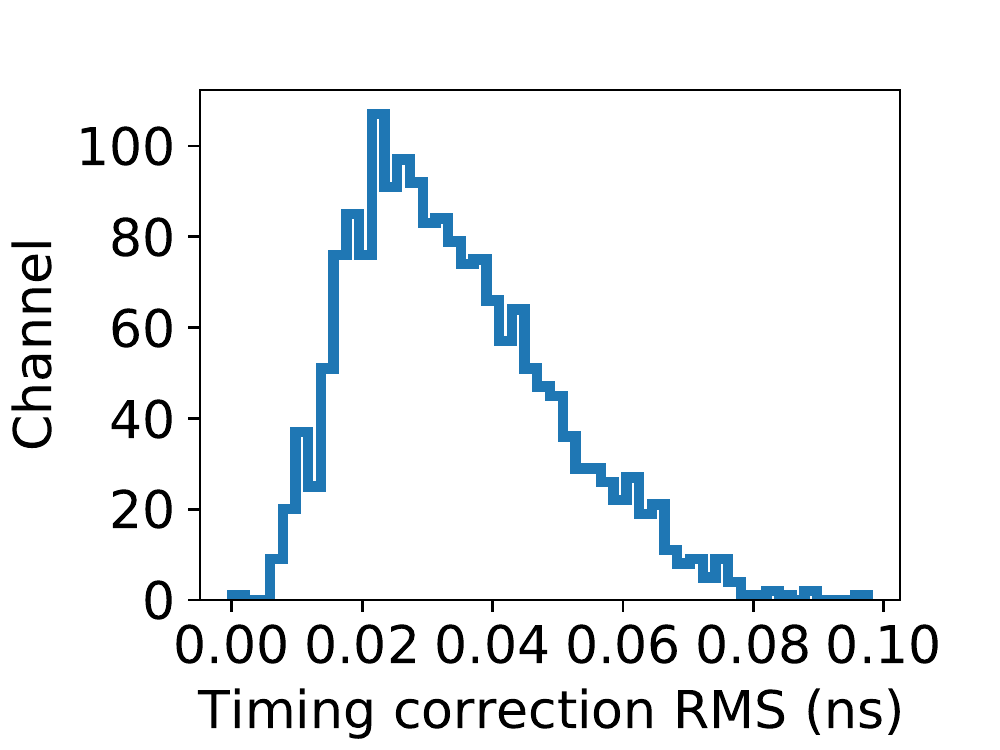}
    \caption{The timing correction coefficients. Left: The timing correction coefficients derived from one example flat-fielding run (performed on 2020-11-20). 
    Middle: The RMS of the timing correction coefficients from the same example flat-fielding run.
    Right: The RMS of the timing correction coefficients over the one year of operation.}
    % run 163816
    \label{fig:timing}
\end{figure}

The distance between the camera and the flat-fielding unit is roughly \SI{36}{\m}, 
and the maximal distance from the pixel at the corners to the center of the camera is roughly \SI{1.15}{\m}.
For the flat-fielding runs, the arrival time of the light pulses across the camera differs by \SI{0.06}{\ns} only, which is negligible.
The timing correction coefficient is defined as the difference between the trigger time of
each pixel and the median trigger time.
The timing correction spreads in the $\pm$ \SI{0.5}{\ns} range, with a standard deviation of no more than \SI{0.25}{\ns} in one flat-fielding run.
The variation of the coefficients per pixel within one year is less than \SI{0.1}{\ns}.
The results are shown in Fig. \ref{fig:timing}.
\subsection{Trigger system}
In the camera trigger system, the PDP is logically mapped onto 588 non-overlapping patches. 
Each patch consists of three neighboring pixels.
The grouping of three neighboring three-pixel patches forms one trigger patch.
The sum of each nine-pixel patch is continuously evaluated by the trigger system,
which is used for computing a trigger decision.
%The setup of the trigger threshold are listed in Table \ref{tab:threshold}.
To reduce the impact of Poisson fluctuations of a few bright pixels on the trigger, 
pixels above a certain NSB rate ("NSB limit" in Table \ref{tab:threshold}.) are ignored while forming the trigger sums.
This significantly improves the operational stability. 
% \textcolor{red}{how many pixels are ignored typically?}
\begin{table}[htbp]
    \centering
    \caption{The setup of the trigger system}
    \label{tab:threshold}
    \begin{tabular}{c|c|c|c}\hline
         Source region                & Run type             & Trigger threshold (9-pixel sum) & NSB limit \\\hline
         \multirow{2}*{Normal}        & ObservationRun          & 69 p.e.                    & \SI{1.1}{\giga\hertz}\\
                                      & MoonlightObservationRun & 104 p.e.                   & \SI{2.7}{\giga\hertz}\\\hline
         \multirow{2}*{Bright}\tablefootnote{Currently: Eta Carinae is set as the only permanent bright region.}
                                      & ObservationRun          & 91 p.e.                    & \SI{2.7}{\giga\hertz}\\
                                      & MoonlightObservationRun & 120 p.e.                   & \SI{2.7}{\giga\hertz}\\\hline
    \end{tabular}
\end{table}

\begin{figure}[t] %[htbp]
    \centering
    \includegraphics[scale=0.3]{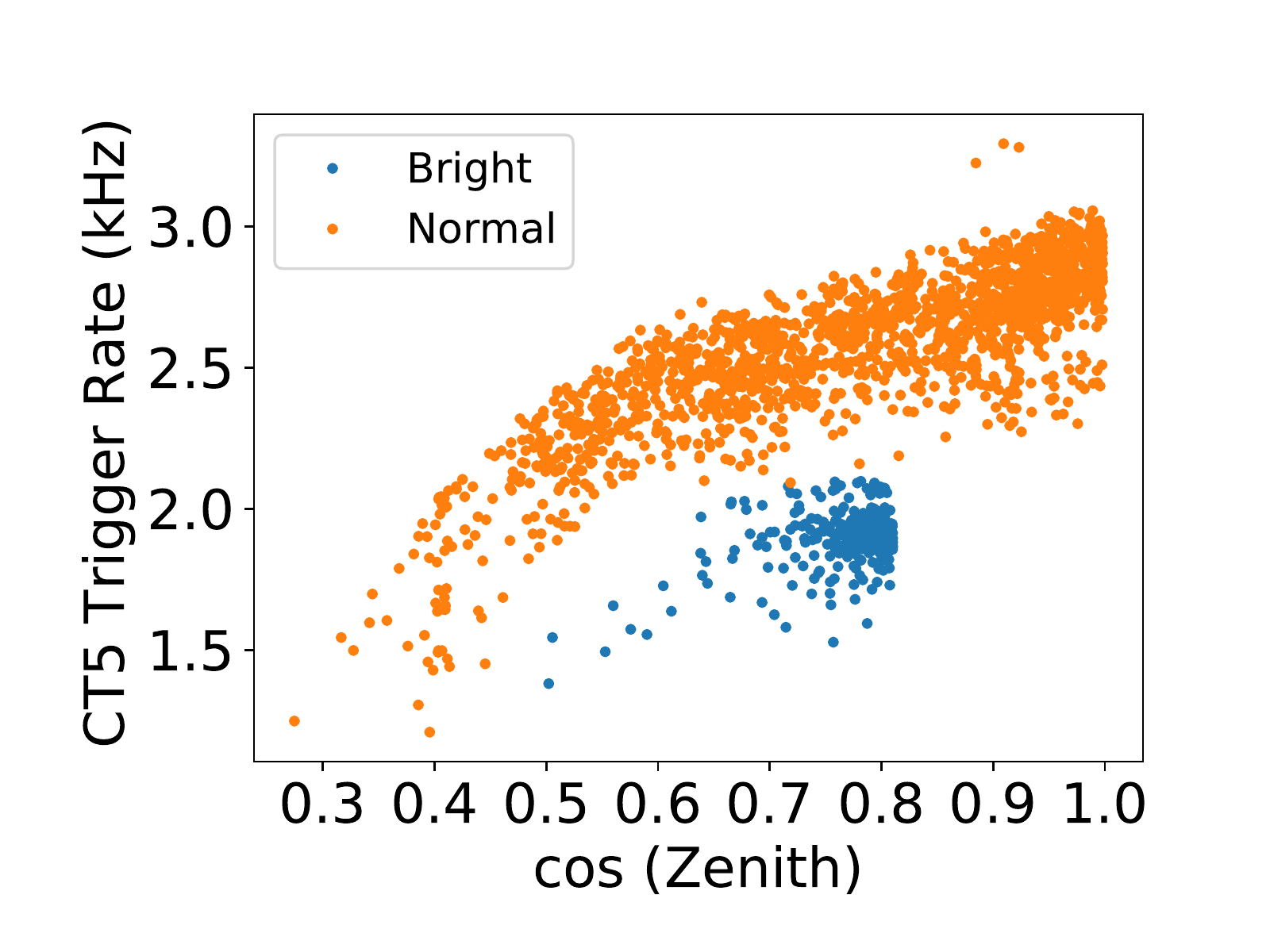}
    \includegraphics[scale=0.3]{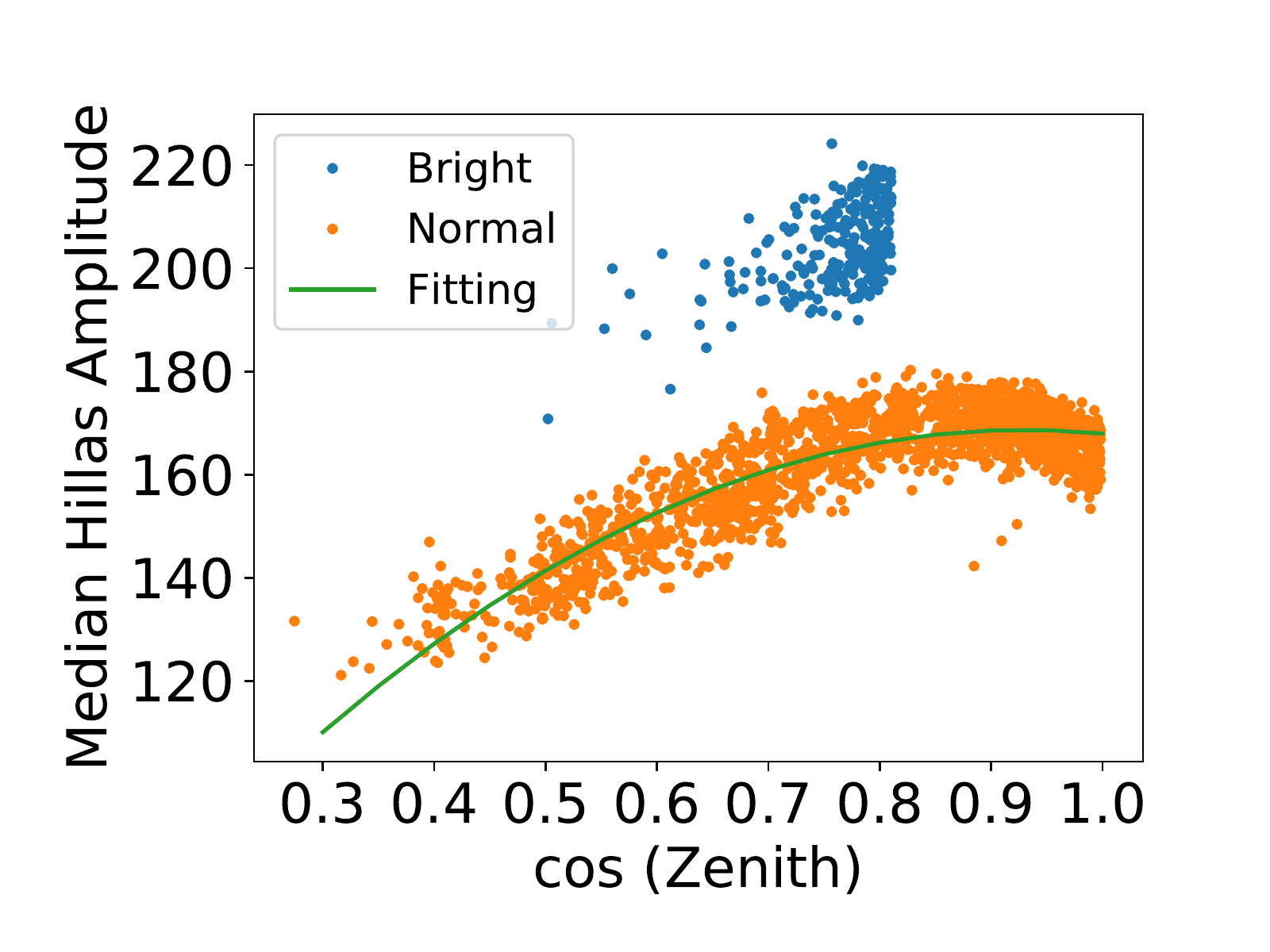}
    \includegraphics[scale=0.3]{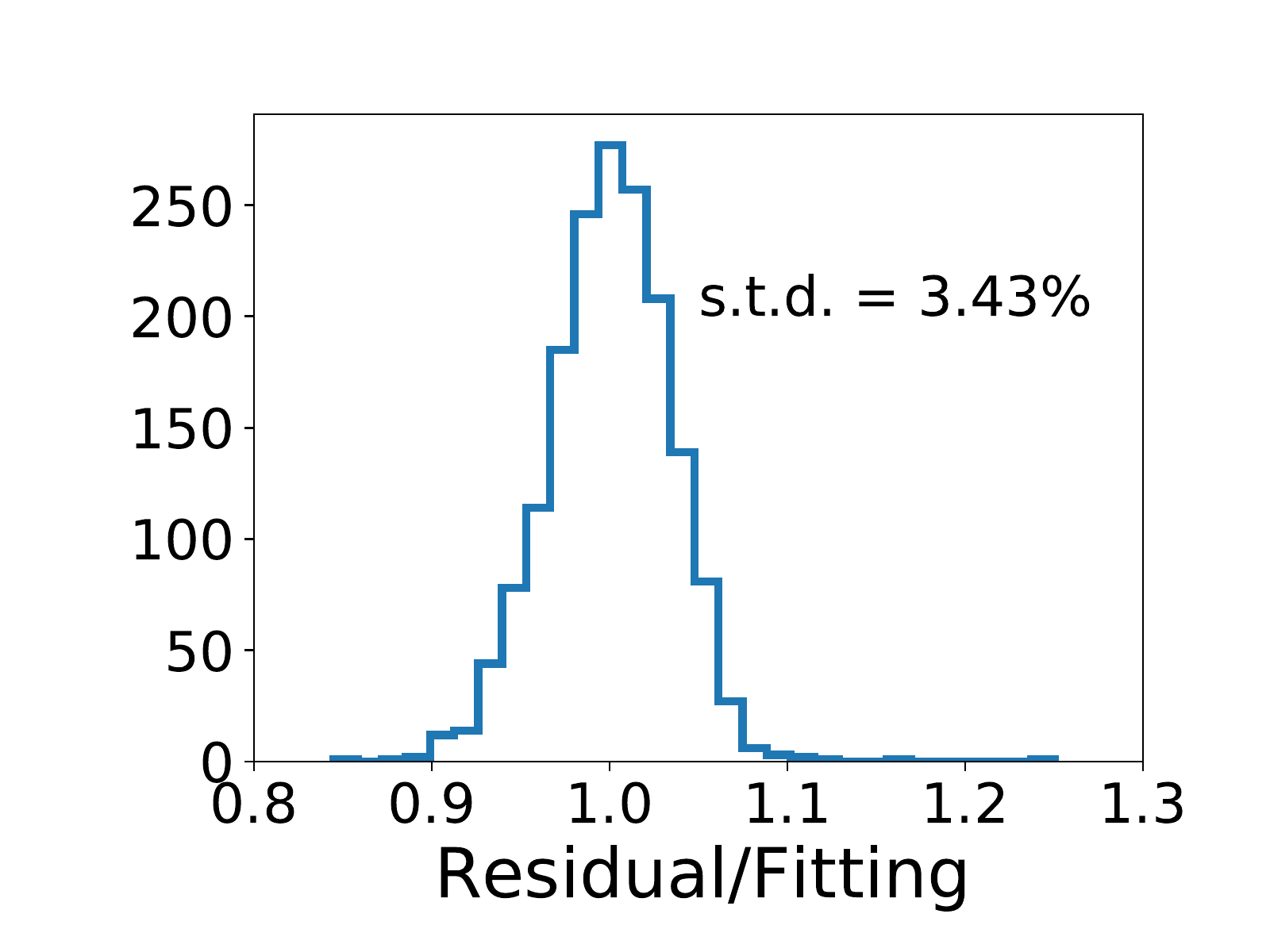}
    \caption{Left: Zenith angle dependency of the trigger rate. Middle: Zenith angle dependency of the Hillas amplitude of all triggered events, fitted with a quadratic function.
    Right: The distribution of the median Hillas amplitudes divided by the quadratic function.}
    \label{fig:trigger}
\end{figure}

The CT5 trigger rates have shown good stability for all observations in one year. 
The average dead time during observation runs is significantly below \SI{0.1}{\percent},
and is caused by backend software limitations.
The runwise-averaged CT5 trigger rates\footnote{Trigger rates are much lower than \SI{30}{\kilo\hertz}, the limitation being the H.E.S.S. readout system.}
show a clear tendency with observation zenith angle, as expected.
Similarly, median Hillas amplitudes of CT5 observation runs vary with the zenith angle, as shown in Fig. \ref{fig:trigger}.  
After correction with a quadratic function of the cosine of the zenith, 
the standard deviation of the median Hillas amplitudes is only \SI{3.43}{\percent} over one year.
This demonstrates the excellent stability of the trigger system relative to the signal reconstruction. \par
To explore the low energy potential of the advanced FlashCam prototype in the CT5, a trigger threshold setting reduced by $\sim$ \SI{15}{\percent} compared to the nominal one of $69\,$p.e. per 9-pixel sum (see Table \ref{tab:threshold}) is used experimentally for dedicated sources, where the energy range of tens of GeV is of increased importance, e.g. pulsars. The Vela sky field was observed successfully with this reduced trigger threshold setting. 

\section{Summary}
Since it was installed into CT5 in November 2019, 
the advanced FlashCam prototype has run smoothly in CT5 for more than one and a half years.
The performance of the camera was stable and excellent.
The scientific verification (SV) observations have been done.
Several targets, e.g. the Crab Nebula, PKS 2155-304, and the Vela pulsar 
have been observed to verify the scientific performance of FlashCam in CT5.
More details are reported in the scientific verification proceeding contribution \href{https://indico.desy.de/event/27991/contributions/102137/}{\#1369} at this conference.

\section*{Acknowledgement}
We acknowledge the excellent work of the technical support staff at our institutes. We also thank the H.E.S.S. collaboration for its support.
%The support of the Namibian authorities and of the University of Namibia in facilitating the construction and operation of H.E.S.S. is gratefully acknowledged, 
%as is the support by the German Ministry for Education and Research (BMBF), the Max Planck Society, 
%the German Research Foundation (DFG), the Helmholtz Association, the Alexander von Humboldt Foundation, 
%the French Ministry of Higher Education, Research and Innovation, 
%the Centre National de la Recherche Scientifique (CNRS/IN2P3 and CNRS/INSU), 
%the Commissariat \`a l'\'energie atomique et aux \'energies alternatives (CEA), 
%the U.K. Science and Technology Facilities Council (STFC), the Knut and Alice Wallenberg Foundation, 
%the National Science Centre, Poland grant no. \\
%2016/22/M/ST9/00382, 
%the South African Department of Science and Technology and National Research Foundation, the University of Namibia, 
%the National Commission on Research, Science \& Technology of Namibia (NCRST), the Austrian Federal Ministry of Education, 
%Science and Research and the Austrian Science Fund (FWF), the Australian Research Council (ARC), 
%the Japan Society for the Promotion of Science and by the University of Amsterdam.

%We appreciate the excellent work of the technical support staff in Berlin, Zeuthen, Heidelberg, Palaiseau, Paris, Saclay, T\"ubingen and in Namibia in the construction and operation of the equipment. This work benefited from services provided by the H.E.S.S. Virtual Organisation, supported by the national resource providers of the EGI Federation.

\end{document}